\documentclass[12pt]{article}
\usepackage{amsmath}
\usepackage[unicode,bookmarks,bookmarksopen,bookmarksopenlevel=2,colorlinks,linkcolor=blue,citecolor=green]{hyperref}

\def\comment#1{}
\input{tcilatex}

\begin{document}

\title{Modified dispersionless Veselov--Novikov equations and corresponding
hydrodynamic chains}
\author{Maxim V. Pavlov \\
%EndAName
P.N. Lebedev Physical Institute, Moscow, Russia}
\date{}
\maketitle

\begin{abstract}
Various links connecting well-known hydrodynamic chains and corresponding
2+1 nonlinear equations are described.
\end{abstract}

\tableofcontents

\textit{keywords}: Riemann mapping, hydrodynamic chain, Gibbons equation

MSC: 35L40, 35L65, 37K10;\qquad PACS: 02.30.J, 11.10.E.

\section{Introduction}

Theory of integrable hydrodynamic chains started with famous paper \textbf{%
\cite{Benney}} (see also, \textbf{\cite{Dubr}}, \textbf{\cite{Gibbons}}, 
\textbf{\cite{Gib+Kod}}, \textbf{\cite{Gib+Tsar}}, \textbf{\cite{Kod+water}}%
, \textbf{\cite{Krich}}, \textbf{\cite{KM}}, \textbf{\cite{Zakh}}). Recent
investigations are obtained in \textbf{\cite{Fer+Kar}} and in \textbf{\cite%
{Maks+Egor}} (see also, \textbf{\cite{Fer+Dav}}, \textbf{\cite{FKP}}).

This paper is devoted to a description of various links associated with the
integrable hydrodynamic chain%
\begin{equation}
B_{t^{1}}^{k}=B_{t^{0}}^{k+1}+\frac{1}{2}%
B^{0}B_{t^{0}}^{k}+(k+1)B^{k}B_{t^{0}}^{0}\text{, \ \ \ \ \ \ }k\in \mathbf{Z%
}.  \label{1}
\end{equation}%
which is a particular case of extended Kupershmidt hydrodynamic chains (see 
\textbf{\cite{Kuper}}, \textbf{\cite{Maks+Kuper}})%
\begin{equation*}
B_{t^{1}}^{k}=B_{t^{0}}^{k+1}+\frac{1}{\beta }B^{0}B_{t^{0}}^{k}+(k+\gamma
)B^{k}B_{t^{0}}^{0}\text{, \ \ \ \ \ \ }k\in \mathbf{Z}.
\end{equation*}

This hydrodynamic chain can be written in the Hamiltonian form%
\begin{equation*}
B_{t^{1}}^{k}=[(2k+1)B^{k+n}D_{t^{0}}+(2n+1)D_{t^{0}}B^{k+n}]\frac{\delta 
\mathbf{H}_{1}}{\delta B^{n}},
\end{equation*}%
where the Hamiltonian is given by $\mathbf{H}_{1}=\int
[B^{1}/3+(B^{0})^{2}/4]dt^{0}$. The extended Kupershmidt hydrodynamic chains
possess an infinite series of conservation laws. All positive conservation
law densities $h_{n}$ are polynomial with respect to positive moments $B^{k}$%
. Negative conservation law densities $h_{-n}$ are more complicated. For
instance, $h_{-1}=-2\exp (B^{-1}/2)$, $h_{-2}=-B^{-2}\exp (3B^{-1}/2)/3$.
Thus, the above hydrodynamic chain possesses an infinite series of commuting
flows. For instance, two first negative flows are%
\begin{eqnarray}
B_{t^{-1}}^{k} &=&e^{B^{-1}/2}[B_{t^{0}}^{k-1}-kB^{k-1}B_{t^{0}}^{-1}],
\label{2} \\
&&  \notag \\
B_{t^{-2}}^{k} &=&e^{3B^{-1}/2}\left[ B_{t^{0}}^{k-2}+\frac{B^{-2}}{2}%
B_{t^{0}}^{k-1}-\left( \frac{3}{2}kB^{k-1}B^{-2}+(k-1)B^{k-2}\right)
B_{t^{0}}^{-1}-kB^{k-1}B_{t^{0}}^{-2}\right] .  \label{3}
\end{eqnarray}

\section{Dispersionless Veselov--Novikov equations}

The first negative commuting flow (\textbf{\ref{2}}) is invariant (see 
\textbf{\cite{Maks+Kuper}}) under a transformation of independent variables $%
t^{0}\leftrightarrow t^{-1}$ and the substitution%
\begin{equation*}
\tilde{B}^{k}=B^{-k-2}e^{(k+1)B^{-1}}.
\end{equation*}%
All other commuting flows are invariant under a transformation of
independent variables $t^{k-1}\leftrightarrow t^{-k}$. For instance, the
second commuting flow (\textbf{\ref{3}}) transforms into (\textbf{\ref{1}}),
while (\textbf{\ref{1}}) transforms into (\textbf{\ref{3}}).

Let us \textit{combine} these commuting flows, i.e. we have new integrable
hydrodynamic chain%
\begin{eqnarray}
B_{t}^{k} &=&B_{t^{0}}^{k+1}+\frac{B^{0}}{2}%
B_{t^{0}}^{k}+(k+1)B^{k}B_{t^{0}}^{0}  \notag \\
&&  \label{4} \\
&&+e^{3B^{-1}/2}\left[ B_{t^{0}}^{k-2}+\frac{B^{-2}}{2}B_{t^{0}}^{k-1}-%
\left( \frac{3}{2}kB^{k-1}B^{-2}+(k-1)B^{k-2}\right)
B_{t^{0}}^{-1}-kB^{k-1}B_{t^{0}}^{-2}\right]  \notag
\end{eqnarray}%
determined by the \textit{mixed} Hamiltonian $\mathbf{H\equiv H}_{1}+\mathbf{%
H}_{-2}=\int [B^{1}/3+(B^{0})^{2}/4-B^{-2}\exp (3B^{-1}/2)/3]dt^{0}$ (see (%
\textbf{\ref{1}}) and (\textbf{\ref{3}})).

Introducing new field variables%
\begin{equation*}
v=\frac{B^{0}}{2}\text{, \ \ \ \ \ \ }u=e^{B^{-1}/2}\text{, \ \ \ \ \ \ \ }w=%
\frac{B^{-2}}{2}e^{B^{-1}},
\end{equation*}%
two conservation laws (see (\textbf{\ref{2}}))%
\begin{equation*}
\partial _{t^{-1}}\frac{B^{0}}{2}=\partial _{t^{0}}e^{B^{-1}/2}\text{, \ \ \
\ \ \ \ \ }\partial _{t^{-1}}e^{B^{-1}/2}=\partial _{t^{0}}\left( \frac{%
B^{-2}}{2}e^{B^{-1}}\right)
\end{equation*}%
can be written in Egorov's form (see \textbf{\cite{Maks+Tsar}}, \textbf{\cite%
{Maks+Egor}})%
\begin{equation*}
v_{t^{-1}}=u_{t^{0}}\text{, \ \ \ \ \ \ }u_{t^{-1}}=w_{t^{0}}.
\end{equation*}%
These conservation laws together with the conservation law (see (\textbf{\ref%
{4}}))%
\begin{equation*}
u_{t}=(uv)_{t^{0}}+(uw)_{t^{-1}}
\end{equation*}%
determine a dispersionless limit of the remarkable symmetric
Veselov--Novikov equation%
\begin{equation}
\Omega _{t^{0}t^{-1}t}=(\Omega _{t^{0}t^{-1}}\Omega
_{t^{0}t^{0}})_{t^{0}}+(\Omega _{t^{0}t^{-1}}\Omega
_{t^{-1}t^{-1}})_{t^{-1}},  \label{sym}
\end{equation}%
where%
\begin{equation*}
v=\Omega _{t^{0}t^{0}}\text{, \ \ \ \ \ \ \ }u=\Omega _{t^{0}t^{-1}}\text{,
\ \ \ \ \ \ \ }w=\Omega _{t^{-1}t^{-1}}.
\end{equation*}

Another 2+1 quasilinear system of the first order can be derived from the
Kupershmidt hydrodynamic chain (\textbf{\ref{1}}) and its first higher
commuting flow%
\begin{equation*}
B_{t^{2}}^{k}=B_{t^{0}}^{k+2}+\frac{3}{2}B^{0}B_{t^{0}}^{k+1}+\left( \frac{%
B^{1}}{2}+\frac{5(B^{0})^{2}}{8}\right)
B_{t^{0}}^{k}+(k+2)B^{k+1}B_{t^{0}}^{0}+(k+1)B^{k}\left( B_{t^{0}}^{1}+\frac{%
5B^{0}}{2}B_{t^{0}}^{0}\right)
\end{equation*}%
determined by the Hamiltonian $\mathbf{H}_{2}=\int
[B^{2}/5+B^{0}B^{1}/2+5(B^{0})^{3}/24]dt^{0}$.

First two conservation laws of the Kupershmidt hydrodynamic chain (\textbf{%
\ref{1}})%
\begin{equation*}
B_{t^{1}}^{0}=\partial _{t^{0}}\left( B^{1}+\frac{3}{4}(B^{0})^{2}\right) 
\text{, \ \ \ \ \ \ \ }\partial _{t^{1}}\left( B^{1}+\frac{3}{4}%
(B^{0})^{2}\right) =\partial _{t^{0}}\left( B^{2}+2B^{0}B^{1}+\frac{3}{4}%
(B^{0})^{3}\right)
\end{equation*}%
and first conservation law of its above higher commuting flow%
\begin{equation*}
B_{t^{2}}^{0}=\partial _{t^{0}}\left( B^{2}+\frac{5}{2}B^{0}B^{1}+\frac{25}{%
24}(B^{0})^{3}\right)
\end{equation*}%
can be written together as a dispersionless limit of non-symmetric
Veselov--Novikov equation%
\begin{equation}
\Omega _{t^{1}t^{1}}=\Omega _{t^{0}t^{2}}-\Omega _{t^{0}t^{0}}\Omega
_{t^{0}t^{1}}+\frac{1}{3}\Omega _{t^{0}t^{0}}^{3},  \label{skew}
\end{equation}%
where%
\begin{equation*}
\Omega _{t^{0}t^{1}}=\frac{B^{1}}{2}+\frac{3}{8}(B^{0})^{2}\text{, \ \ \ \ \
\ }\Omega _{t^{0}t^{2}}=\frac{B^{2}}{2}+\frac{5}{4}B^{0}B^{1}+\frac{25}{48}%
(B^{0})^{3}.
\end{equation*}

\textbf{Remark}: The symmetric dispersionless Veselov--Novikov equation (%
\textbf{\ref{sym}}) can be obtained from the compatibility condition $%
\partial _{t^{1}}(\partial _{t^{-1}}p)=\partial _{t^{-1}}(\partial
_{t^{1}}p) $, where the generating function of conservation law densities $p$
satisfies%
\begin{equation*}
p_{t^{-1}}=-\partial _{t^{0}}\frac{u}{p}\text{, \ \ \ \ \ \ \ \ \ }%
p_{t}=\partial _{t^{0}}\left( \frac{p^{3}}{3}+vp-\frac{uw}{p}-\frac{u^{3}}{%
3p^{3}}\right) ,
\end{equation*}%
while the non-symmetric dispersionless Veselov--Novikov equation (\textbf{%
\ref{skew}}) can be obtained from the compatibility condition $\partial
_{t^{1}}(\partial _{t^{2}}p)=\partial _{t^{2}}(\partial _{t^{1}}p)$, where
the generating function of conservation law densities $p$ satisfies%
\begin{equation*}
p_{t^{1}}=\partial _{t^{0}}\left( \frac{p^{3}}{3}+\Omega
_{t^{0}t^{0}}p\right) \text{, \ \ \ \ \ \ }p_{t^{2}}=\partial _{t^{0}}\left[ 
\frac{p^{5}}{5}+\Omega _{t^{0}t^{0}}p^{3}+\left( \Omega _{t^{0}t^{1}}+\Omega
_{t^{0}t^{0}}^{2}\right) p\right] .
\end{equation*}%
Corresponding Gibbons equations (see \textbf{\cite{Gibbons}}, \textbf{\cite%
{Maks+Ham}}) associated with the above hydrodynamic chains are%
\begin{eqnarray*}
\lambda _{t^{1}}-\left( p^{2}+v\right) \lambda _{t^{0}} &=&\frac{\partial
\lambda }{\partial p}\left[ p_{t^{1}}-\partial _{t^{0}}\left( \frac{p^{3}}{3}%
+vp\right) \right] , \\
&& \\
\lambda _{t^{-2}}-\left( \frac{uw}{p^{2}}+\frac{u^{3}}{p^{4}}\right) \lambda
_{t^{0}} &=&\frac{\partial \lambda }{\partial p}\left[ p_{t^{-2}}+\partial
_{t^{0}}\left( \frac{uw}{p}+\frac{u^{3}}{3p^{3}}\right) \right] , \\
&& \\
\lambda _{t}-\left( p^{2}+v+\frac{uw}{p^{2}}+\frac{u^{3}}{p^{4}}\right)
\lambda _{t^{0}} &=&\frac{\partial \lambda }{\partial p}\left[
p_{t}-\partial _{t^{0}}\left( \frac{p^{3}}{3}+vp-\frac{uw}{p}-\frac{u^{3}}{%
3p^{3}}\right) \right] , \\
&& \\
\lambda _{t^{-1}}-\frac{u}{p^{2}}\lambda _{t^{0}} &=&\frac{\partial \lambda 
}{\partial p}\left( p_{t^{-1}}+\partial _{t^{0}}\frac{u}{p}\right) , \\
&& \\
\lambda _{t^{2}}-\left( p^{4}+3vp^{2}+v^{2}+s\right) \lambda _{t^{0}} &=&%
\frac{\partial \lambda }{\partial p}\left[ p_{t^{2}}-\partial _{t^{0}}\left( 
\frac{p^{5}}{5}+vp^{3}+(v^{2}+s)p\right) \right] ,
\end{eqnarray*}%
where $v=\Omega _{t^{0}t^{0}}$, $s=\Omega _{t^{0}t^{1}}$, and (\textbf{\ref%
{skew}}) can be written as 2+1 quasilinear system of the first order%
\begin{equation*}
\partial _{t^{1}}v=\partial _{t^{0}}s\text{, \ \ \ \ }\partial
_{t^{1}}s=\partial _{t^{2}}v+\partial _{t^{0}}\left( \frac{v^{3}}{3}%
-vs\right) .
\end{equation*}

\textbf{Remark}: Since the function $\Omega $ is unique for both
Veselov--Novikov equations, both of them possess the same set of
hydrodynamic reductions determined by the Gibbons--Tsarev system (see 
\textbf{\cite{Gib+Tsar}}, \textbf{\cite{Maks+Kuper}}).

\section{Modified VN equation}

The generating function of conservation laws and commuting flows (see 
\textbf{\cite{Maks+Kuper}}) is given by%
\begin{equation}
\partial _{\tau (\zeta )}p(\lambda )=\frac{1}{2}\partial _{t^{0}}\ln \frac{%
p(\lambda )-p(\zeta )}{p(\lambda )+p(\zeta )},  \label{gen}
\end{equation}%
where $p(\lambda )$ can be found due to the B\"{u}rmann--Lagrange series
(see, for instance, \textbf{\cite{Lavr}}) from the Riemann mappings%
\begin{equation*}
\lambda |_{p\rightarrow \infty }=p^{2}\exp \left( \underset{k=0}{\overset{%
\infty }{\sum }}\frac{B^{k}}{p^{2(k+1)}}\right) \text{, \ \ \ \ \ \ \ }%
\lambda |_{p\rightarrow 0}=p^{-2}\exp \left( \underset{k=1}{\overset{\infty }%
{\sum }}B^{-k}p^{2(k-1)}\right) .
\end{equation*}

In this section we consider another expansion%
\begin{equation}
p(\lambda )=\tilde{h}_{0}+\lambda \tilde{h}_{1}+\lambda ^{2}\tilde{h}_{2}+...%
\text{, \ \ \ \ }\lambda \rightarrow 0,  \label{taylor}
\end{equation}%
where $\lambda $ is another local parameter. In such a case, (\textbf{\ref%
{gen}}) leads to an infinite series of generating functions%
\begin{equation*}
\partial _{\tau ^{0}}p(\lambda )=\frac{1}{2}\partial _{t^{0}}\ln \frac{%
p(\lambda )-\tilde{h}_{0}}{p(\lambda )+\tilde{h}_{0}}\text{, \ \ \ \ \ }%
\partial _{\tau ^{1}}p(\lambda )=-\partial _{t^{0}}\frac{\tilde{h}%
_{1}p(\lambda )}{p^{2}(\lambda )-\tilde{h}_{0}^{2}},...,
\end{equation*}%
where%
\begin{equation*}
\partial _{\tau (\lambda )}=\partial _{\tau ^{0}}+\lambda \partial _{\tau
^{1}}+\lambda ^{2}\partial _{\tau ^{2}}+...
\end{equation*}

Introducing a new generating function of conservation law densities%
\begin{equation}
\tilde{p}(\lambda )=\ln \frac{p(\lambda )+\tilde{h}_{0}}{p(\lambda )-\tilde{h%
}_{0}},  \label{miura}
\end{equation}%
and eliminating $t^{0}$, both above generating functions can be written in
the form%
\begin{equation}
\partial _{\tau ^{1}}\tilde{p}(\lambda )=\partial _{\tau ^{0}}\left( \frac{%
\tilde{h}_{1}}{\tilde{h}_{0}}\sinh \tilde{p}(\lambda )\right)  \label{sin}
\end{equation}%
which is a first coefficient from (\textbf{\ref{gen}}) (see (\textbf{\ref%
{miura}}) and \textbf{\cite{Maks+Egor}})%
\begin{equation*}
\partial _{\tau (\zeta )}\tilde{p}(\lambda )=\partial _{\tau ^{0}}\ln \frac{%
e^{\tilde{p}(\lambda )}-e^{\tilde{p}(\zeta )}}{e^{\tilde{p}(\lambda )+\tilde{%
p}(\zeta )}-1}.
\end{equation*}

Substituting the Taylor expansion (\textbf{\ref{taylor}}) into the above
generating functions, one can obtain new 2+1 nonlinear equations%
\begin{equation*}
\Omega _{\tau ^{0}\tau ^{0}}=\frac{1}{2}\ln \frac{\Omega _{t^{0}\tau ^{1}}}{%
\Omega _{t^{0}\tau ^{0}}}\text{, \ \ \ \ \ \ }\Omega _{\tau ^{1}\tau
^{1}}=\Omega _{\tau ^{0}\tau ^{2}}-\Omega _{\tau ^{0}\tau ^{1}}^{2}+\frac{1}{%
2}\exp (4\Omega _{\tau ^{0}\tau ^{0}}),
\end{equation*}%
where $\tilde{h}_{k}\equiv \Omega _{t^{0}\tau ^{k}}$, because $p(\lambda
)\equiv \Omega _{t^{0}\tau (\lambda )}$ (the last formula is a consequence
of (\textbf{\ref{gen}}), if $p\rightarrow \infty $).

\textbf{Remark}: The second above equation was found in \textbf{\cite%
{Maks+Egor}}. This is a nontrivial generalization of a continuum limit of
discrete KP hierarchy (see, for instance, \textbf{\cite{Lei Yu}}). Thus, we
described links between different well-known 2+1 integrable equations in
this paper.

\textbf{Theorem}: \textit{The hydrodynamic chain}%
\begin{equation}
A_{\tau ^{1}}^{k}=\frac{A_{\tau ^{0}}^{k+1}}{A^{0}}-\frac{%
(k+2)A^{k+1}-kA^{k-1}}{(A^{0})^{2}}A_{\tau ^{0}}^{0}\text{, \ \ \ \ \ }%
k=0,1,2,...  \label{one}
\end{equation}%
\textit{satisfies the Gibbons equation}%
\begin{equation}
\lambda _{\tau ^{1}}-\frac{\cosh \tilde{p}}{A^{0}}\cdot \lambda _{\tau
^{0}}=\lambda _{\tilde{p}}\left( \tilde{p}_{\tau ^{1}}-\partial _{\tau ^{0}}%
\frac{\sinh \tilde{p}}{A^{0}}\right) ,  \label{two}
\end{equation}%
\textit{where the equation of the Riemann surface is given by}%
\begin{equation}
\lambda =\underset{n=0}{\overset{\infty }{\sum }}\frac{A^{n}}{\cosh ^{n+1}%
\tilde{p}}.  \label{three}
\end{equation}

\textbf{Proof}: The substitution (\textbf{\ref{one}}) and (\textbf{\ref%
{three}}) in (\textbf{\ref{two}}) leads to an identity.

\textbf{Theorem}: \textit{The above hydrodynamic chain }(\textbf{\ref{one}}) 
\textit{is Hamiltonian}%
\begin{equation*}
A_{\tau ^{1}}^{k}=\left[ [(k+2)A^{k+1}-kA^{k-1}]D_{\tau ^{0}}+A_{\tau
^{0}}^{k+1}\right] \frac{\partial h}{\partial A^{0}},
\end{equation*}%
\textit{where the Hamiltonian density }$h=\ln A^{0}$.

\textbf{Proof}: Integrable hydrodynamic chains associated with more general
Hamiltonian structure were considered in \textbf{\cite{Maks+algebr}}.

\textbf{Remark}: Simplest hydrodynamic reductions%
\begin{equation*}
a_{\tau ^{1}}^{i}=\partial _{\tau ^{0}}\frac{\sinh a^{i}}{A^{0}}
\end{equation*}%
are given by the moment decomposition%
\begin{equation*}
A^{k}=\frac{1}{k+1}\underset{i=1}{\overset{N}{\sum }}\epsilon _{i}(\cosh
a^{i})^{k+1}\text{, \ \ \ \ \ }k=0,1,2,...\text{, \ \ \ \ \ \ }\underset{i=1}%
{\overset{N}{\sum }}\epsilon _{i}=0.
\end{equation*}

\textbf{Theorem}: 2+1 \textit{nonlinear equation}%
\begin{equation*}
\Omega _{\tau ^{1}\tau ^{1}}=\Omega _{\tau ^{0}\tau ^{2}}-\frac{1}{2}\Omega
_{\tau ^{0}\tau ^{1}}^{2}+\epsilon \exp (-2\Omega _{\tau ^{0}\tau ^{0}})
\end{equation*}%
\textit{can be determined from the compatibility condition }$\partial _{\tau
^{1}}(\partial _{\tau ^{2}}p)=\partial _{\tau ^{2}}(\partial _{\tau ^{1}}p)$%
, \textit{where the generating functions of conservation laws are given by}%
\begin{equation*}
p_{\tau ^{1}}=\partial _{\tau ^{0}}[e^{-U}(e^{p}+\epsilon e^{-p})]\text{, \
\ \ \ \ \ }p_{\tau ^{2}}=\partial _{\tau ^{0}}\left[ \frac{e^{2p}-\epsilon
^{2}e^{-2p}}{2}e^{-2U}-V(e^{p}-\epsilon e^{-p})e^{-U}\right] ,
\end{equation*}%
\textit{where }$U=\Omega _{\tau ^{0}\tau ^{0}}$, $V=\Omega _{\tau ^{0}\tau
^{1}}$.

\textbf{Remark}: Substituting the transformation%
\begin{equation*}
\mu =e^{-U}(e^{p}+\epsilon e^{-p})-V
\end{equation*}%
into the above generating functions and eliminating $\tau ^{0}$, one can
obtain the generating function of conservation laws%
\begin{equation*}
\mu _{\tau ^{2}}=\partial _{\tau ^{1}}\left[ \frac{\mu +V}{2}\sqrt{(\mu
+V)^{2}-4\epsilon e^{-2U}}-V(\mu +V)-W\right]
\end{equation*}%
generalizing ($\epsilon =0$) the generating function of conservation laws
for the Benney hydrodynamic chain%
\begin{equation*}
\mu _{\tau ^{2}}=\partial _{\tau ^{1}}\left[ \frac{\mu ^{2}}{2}-\frac{V^{2}}{%
2}-W\right] ,
\end{equation*}%
where $W=\Omega _{\tau ^{0}\tau ^{2}}$. Such a transformation for $\epsilon
=0$ was found first by Yuji Kodama (see, for instance, \textbf{\cite{Lei Yu}}%
).

Let us introduce new notations $U=-C^{-1}$, $V=-C^{0}$ and $q=p-U$.

\textbf{Theorem}: \textit{The integrable hydrodynamic chain}%
\begin{eqnarray}
C_{\tau ^{1}}^{-1} &=&C_{\tau ^{0}}^{0}\text{, \ \ \ \ \ \ }C_{\tau
^{1}}^{0}=\partial _{\tau ^{0}}\left( C^{1}-\epsilon e^{-2C^{-1}}\right) , 
\notag \\
&&  \label{5} \\
C_{\tau ^{1}}^{k} &=&C_{\tau ^{0}}^{k+1}+kC^{k}C_{\tau ^{0}}^{0}-\epsilon
e^{-2C^{-1}}[C_{\tau ^{0}}^{k-1}+2(k-1)C^{k-1}C_{\tau ^{0}}^{-1}]\text{, \ \
\ \ \ \ }k=1,2,...  \notag
\end{eqnarray}%
\textit{satisfies the Gibbons equation (see} (\textbf{\ref{sin}}))%
\begin{equation}
\lambda _{\tau ^{1}}-(e^{q}-\epsilon e^{-2C^{-1}-q})\lambda _{\tau ^{0}}=%
\frac{\partial \lambda }{\partial q}\left[ q_{\tau ^{1}}-\partial _{\tau
^{0}}\left( e^{q}+C^{0}+\epsilon e^{-2C^{-1}-q}\right) \right] ,  \label{gib}
\end{equation}%
\textit{where the Riemann mapping is determined by}%
\begin{equation}
\lambda =e^{q}+\underset{n=0}{\overset{\infty }{\sum }}C^{n}e^{-nq}.
\label{exp}
\end{equation}

\textbf{Proof}: Indeed, a substitution the above series and the hydrodynamic
chain (\textbf{\ref{5}}) into the Gibbons equation (\textbf{\ref{gib}})
leads to an identity.

This Riemann mapping (\textbf{\ref{exp}}) is the same as for a continuum
limit of two-dimensional Toda lattice. This hydrodynamic chain (\textbf{\ref%
{5}}) is a member of an integrable hierarchy generalizing ($\epsilon =0$) an
integrable hierarchy containing a continuum limit of two-dimensional Toda
lattice.

\section*{Acknowledgement}

I thank Leonid Bogdanov, Eugeni Ferapontov and Boris Kupershmidt for their
stimulating and clarifying discussions.

I am grateful to the Institute of Mathematics in Taipei (Taiwan) where some
part of this work has been done, and especially to Jen-Hsu Chang, Jyh-Hao
Lee, Ming-Hien Tu and Derchyi Wu for fruitful discussions.

\end{document}